\documentclass[showpacs,twocolumn, numbers,amsmath,amssymb,prl]{revtex4-1}
\usepackage{graphicx}
\usepackage{amsmath, amsfonts, amsthm, amssymb,mathtools}
\usepackage{latexsym}
\usepackage{bm}
\usepackage{ulem}
\usepackage{dcolumn}
\usepackage{epstopdf}

\usepackage{color}

\newcommand{\Pt}{$^{195}$Pt}

\newcommand{\upc}{U$_2$PtC$_2$}

\newcommand{\Hparac}{$\mathbf{H}_0\parallel\hat{c}$}

\newcommand{\Hperpc}{$\mathbf{H}_0\perp \hat{c}$}

\begin{document}

\title{Evidence for spin-triplet superconductivity in \upc\ from \Pt\ NMR}
\author{A.M. Mounce}

\author{H. Yasuoka}

\author{G. Koutroulakis}
\altaffiliation{Department of Physics $\&$ Astronomy, UCLA, Los Angeles, CA 90095, USA}

\author{N. Ni}
\altaffiliation{Department of Physics $\&$ Astronomy, UCLA, Los Angeles, CA 90095, USA}

\author{E.D. Bauer}

\author{F. Ronning}

\author{J.D. Thompson}
\affiliation{Los Alamos National Laboratory, Los Alamos, NM 87545, USA}

\date{Version \today}

\pacs{}

\begin{abstract}
Nuclear magnetic resonance (NMR) measurements on the $^{195}$Pt nucleus in an aligned powder of the moderately heavy-fermion material U$_2$PtC$_2$ are consistent with spin-triplet pairing in its superconducting state.   Across the superconducting transition temperature and to much lower temperatures, the NMR Knight shift is temperature independent for field  both parallel and perpendicular to the tetragonal $c$-axis, expected for triplet equal-spin pairing superconductivity. The NMR spin-lattice relaxation rate $1/T_1$, in the normal state, exhibits characteristics of ferromagnetic fluctuations, compatible with an enhanced Wilson ratio. In the superconducting state, $1/T_1$ follows a power law with temperature without a coherence peak giving additional support that \upc\ is an unconventional superconductor. Bulk measurements of the AC-susceptibility and resistivity indicate that the upper critical field exceeds the Pauli limiting field for spin-singlet pairing and is near the orbital limiting field, an additional indication for spin-triplet pairing. 

\end{abstract}

\maketitle

The competition between the localized and itinerant nature of the 5$f$ electrons in heavy fermion uranium superconductors leads to a rich variety of electronic and magnetic properties in these materials.  One such uranium compound \upc , discovered 45 years ago \cite{mat69}, has been known to be superconducting with a transition temperature $T_c$ = 1.47 K, without magnetic order.  A general measure of the local or itinerant nature of $f$ electron materials is  based on the inter-atomic spacing and, for uranium, this distance is $d_{U-U} \approx$ 3.6 \AA \cite{hil70}. Below this limit, $f$-bands overlap, electrons tend to be itinerant, and phonon-mediated superconductivity can be supported, as for example in U$_6$Fe \cite{del83}. Above this limit, however, $f$ electrons tend to be more localized and electronic correlations in narrow bands of hybridized $f$ and ligand states become important, possibly leading to unconventional superconductivity as is the case for the heavy-fermion UPt$_3$ \cite{ste84}. \upc\, with $d_{U-U} = $ 3.52 \AA\ is intermediate to these extremes \cite{mei84}, leaving to question the nature of superconductivity in this material.  In fact, early experiments of muon-spin resonance \cite{wu94} and pressure-dependent resistivity\cite{tho85} have shown initial indications of an unconventional superconducting state.  Nevertheless, the microscopic details of the superconducting and electronic properties are largely unknown to date, likely due to the difficulty of making high quality samples.  

Here, we report the first $^{195}$Pt NMR measurements of \upc\ to determine the microscopic details of the normal and superconducting (SC) states.  We find evidence for strong ferromagnetic fluctuations in the normal state from a modified Korringa law and Wilson ratio.  Additionally, the NMR shift is temperature independent in the SC state and, from bulk measurements, the upper critical field is much larger than the BCS Pauli limiting field for spin-singlet superconductivity, both consistent with spin-triplet superconductivity.  Superconductivity near ferromagnetism has been found in only a few uranium compounds, namely UGe$_2$ (SC under pressure) \cite{sax00}, URhGe \cite{aok01}, and UCoGe \cite{huy07}, all with $d_{U-U} \approx$ 3.5 \AA . These materials order ferromagnetically in addition to their superconductivity, and have an upper critical field that exceeds the Pauli limiting value \cite{she01, aok01, huy07}; whereas, UCoGe also has a temperature-independent spin shift, $K_s$, in its SC state \cite{hat14}. \upc \  is similar to these materials with regards to the temperature-independent shift and upper critical field, although it is distinct in that it has strong ferromagnetic fluctuations rather than static ferromagnetic order.

\begin{figure}[b!]
\begin{center}
		\includegraphics[width=.45\textwidth]{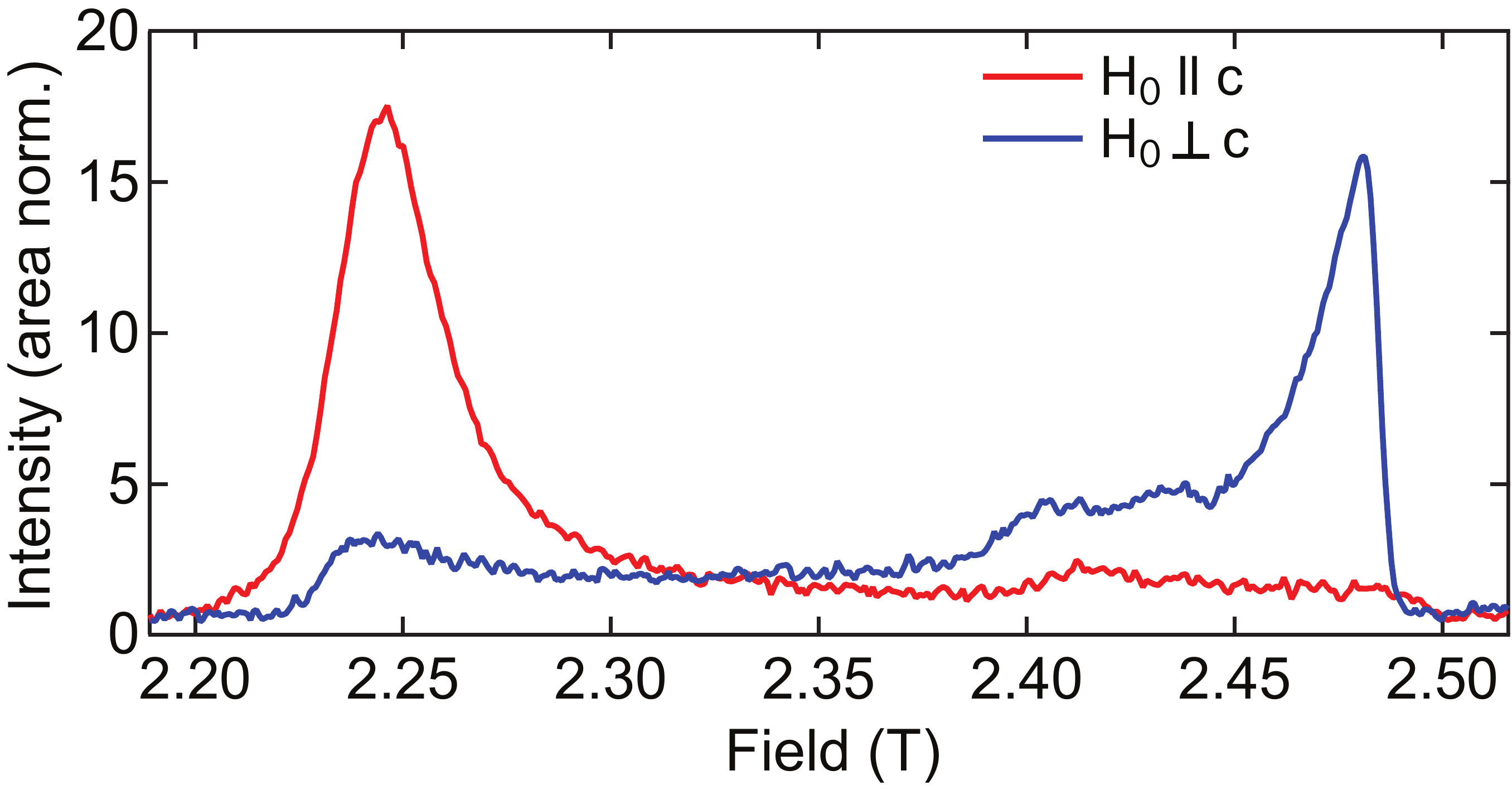}
		\caption{The \Pt\ NMR field swept aligned powder spectra for \Hparac\ (red) and \Hperpc\ (blue) normalized by the spectral area taken at a carrier frequency of $\omega_0 \approx 23$ MHz, at a temperature $T = 1.8 $ K.  The \upc\ aligned powder spectra show an angular dependence consistent with an anisotropic NMR shift and successful alignment.}
\label{spectra}
\end{center}
\end{figure}

Our polycrystalline samples were arc melted then annealed for approximately one month resulting in a  $RRR\approx$ 30 with a residual resistivity of $\rho_0\approx 10\ \mu \Omega$cm. To perform NMR experiments, a polycrystalline sample was  powdered, mixed with stycast 1266 epoxy in a 1:1 ratio, poured into a teflon cup, and aligned along a radial direction of the cup in a field of 9 T for 24 hours until the epoxy cured. The sample used for NMR had a \upc\ mass of $m\approx 70$ mg. The $^{195}$Pt NMR spectra were taken by the field sweep method, at fixed resonant frequency, using a typical $\pi/2 - \tau - \pi$ Hahn echo pulse sequence.  \Pt\ spin-lattice relaxation time ($T_1$) measurements were taken by an inversion recovery sequence at the peak of the spectrum for both orientations. AC-susceptibility measurements were taken \textit{in situ} by recording the detuning of the NMR resonant circuit as a function of magnetic field and temperature, from which $H_{c2}$ and $T_c$ were deduced respectively.

With the external field along the alignment axis, \Hparac, and perpendicular to the alignment axis, \Hperpc ,  at a carrier frequency $\omega_0 \approx$ 23 MHz, there are well defined peaks for each orientation indicating successful alignment, as shown in Fig. \ref{spectra}. There are minor components with small spectral weight which retain a powder like pattern indicating a small fraction of misaligned crystallites. There are also small features near the peak for \Hperpc\ that are  orientation independent, and which may be associated with a secondary phase.  Despite these slight imperfections in the spectra, NMR is selective in that measurements of the shift, $K$, or $T_1$ can be made at the peaks of the spectra for either orientation and, therefore, are independent of misaligned or impurity components.

\begin{figure}[t!]
\begin{center}
		\includegraphics[width=.45\textwidth]{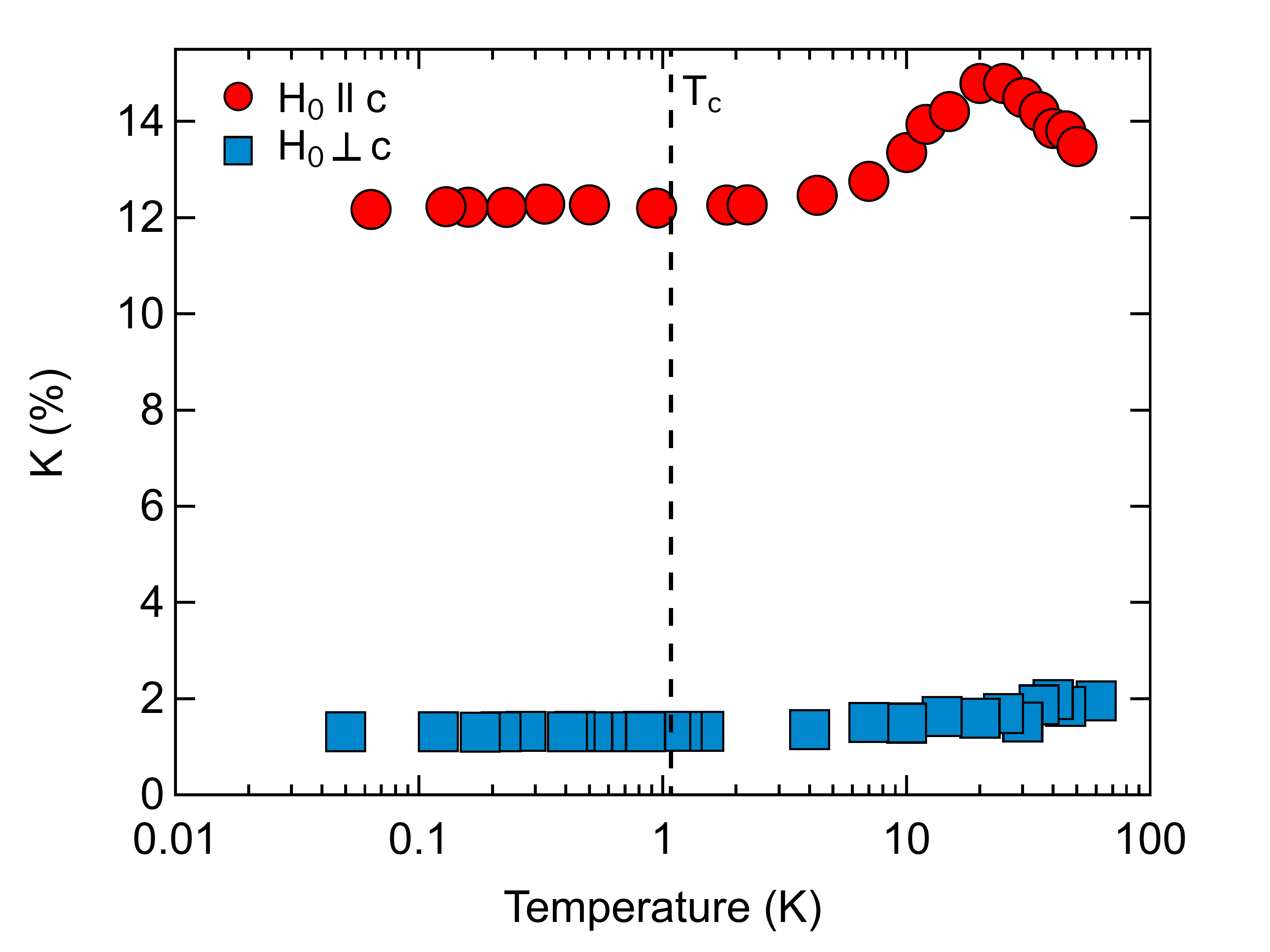}
		\caption{The NMR shift in both the normal and superconducting states for \Hparac\ and \Hperpc .  In the normal state, the shifts for both orientations follow the bulk susceptibility down to a characteristic temperature $T^*=$ 25 K. At temperatures $T^*> T > T_c$ the shift deviates from the bulk susceptibility, and in the superconducting state, $T_c(H_0) \approx 1.1 $K, the shift becomes temperature independent.}
\label{shifts}
\end{center}
\end{figure}

For a field swept spectrum, the NMR shift, $K$, can be expressed as the resonance field relative to that expected for the bare gyromagnetic ratio of the nucleus, $^{195}\gamma/2\pi $= 9.171 MHz/T, at fixed frequency $\omega_0$, as $K = \frac{\omega_0 / \gamma-H_{res}}{H_{res}}$, where $H_{res}$ is the resonant field of maximum spectral intensity. The shift is the sum of components such that $K = K_s(T) + K_o$, where $K_s = A\chi_s(T)$ is the spin shift equal to the hyperfine  coupling constant $A$ times the temperature-dependent electronic spin susceptibility at the Fermi surface $\chi_s(T)$, and $K_o$ is the orbital shift which is proportional to the temperature independent Van Vleck susceptibility.   As shown in Fig. \ref{shifts}, in the normal state, the shift for \Hparac\ increases from $T =$ 50 K down to a characteristic temperature $T^* = 25$ K, below which it decreases down to $T_c$. For \Hperpc, the shift monotonically decreases from high temperature down to  $T_c$. For both field directions, there is no detectable change in the shift below $T_c$.

\begin{figure}[b!]
\begin{center}
		\includegraphics[width=.45\textwidth]{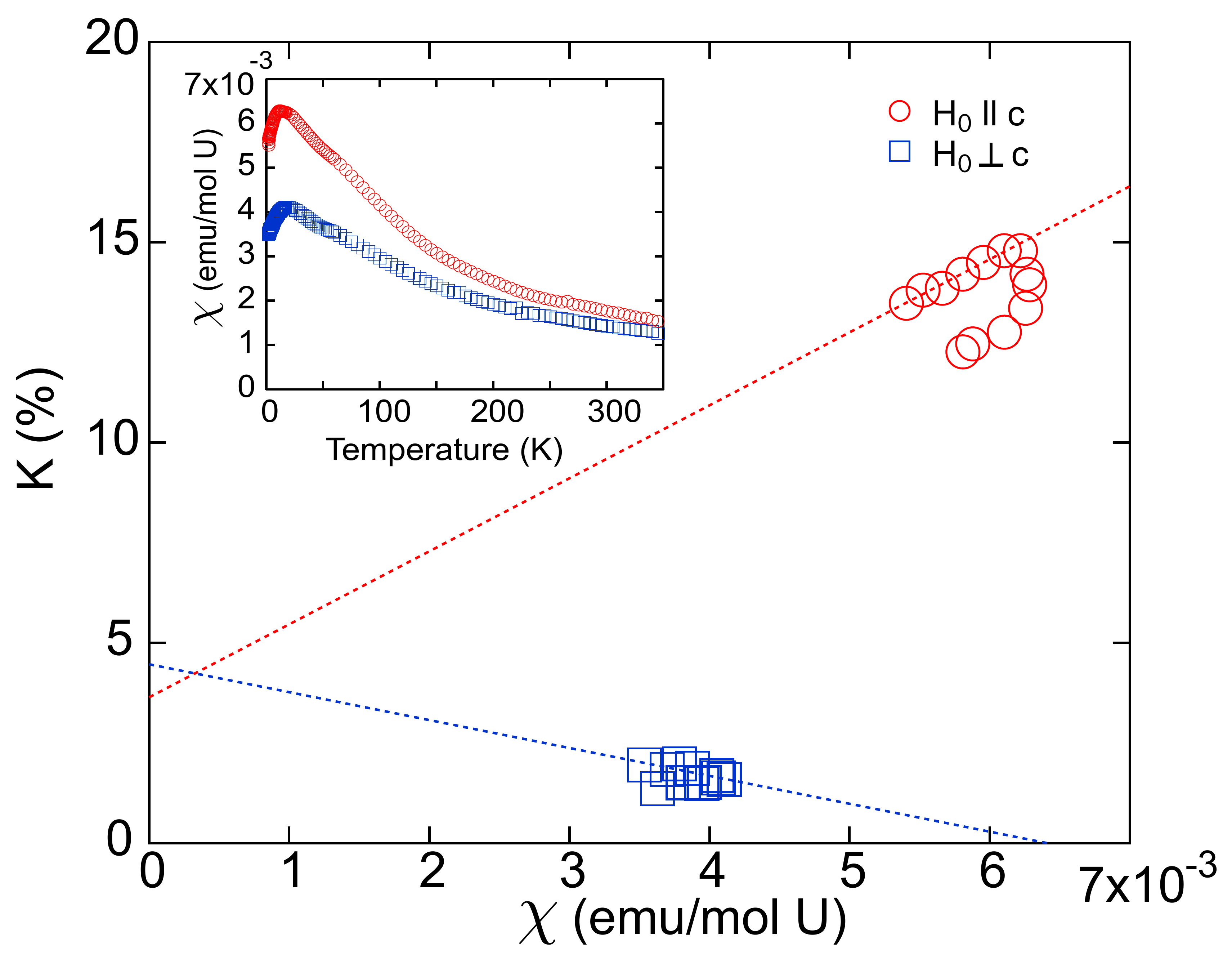}
		\caption{The $K$-$\chi$ plot for both \Hparac\ and \Hperpc . (Inset) The bulk susceptibility measured by SQUID magnetometry for both orientations. $\chi$ was determined from the difference in magnetization measured in fields of 4 and 5T.}
\label{kchi}
\end{center}
\end{figure}

By plotting $K(T)$ vs the bulk susceptibility $\chi(T)$, Fig. \ref{kchi}, with temperature as an implicit parameter, $A$ and $K_o$ can be found by a linear fit, as the slope and the $y$-intercept respectively.  Here, the $K-\chi$ plot deviates from linearity for $T<T^* = 25$ K for both orientations.  This Knight shift anomaly is a common feature among the heavy fermion materials \cite{cur09}. Using data from the temperature region where $K\propto \chi$, we find $A_c$ =101.2 $\pm$ 6.9 kOe/$\mu_B$ and $K_{o,c}$ = 3.64 $\pm$ 0.72 \% for \Hparac, while for \Hperpc ,  $A_{ab} = -38.9 \pm 14.0 $ kOe/$\mu_B$ and $K_{o,ab} = 4.46$ $\pm$ 0.97 \%.  

In the SC state, for a singlet superconductor in the clean low-field limit, $K_s$ should decrease below $T_c$ down to the value of $K_o$ as $T \to 0$ K.  For \upc, below $T_c$ and along both orientations, $K_s$ is temperature independent, Fig. \ref{shifts}. However, it is important to consider how much the shift does not change relative to the expected behavior of a singlet superconductor. By considering the total shift $K$ and the orbital shift  $K_o$ found from the normal state, the magnitude of the spin shifts in the SC state are $K_{s,c}(T\leq T_c) = 8.63$ \% and $K_{s,ab}(T\leq T_c) = -3.15$ \%.  These expected values of $K_s$ are much greater than the line width of the spectra accounting for an uncertainty of less than 0.2 \%. The temperature independence of the spin component of the shift below $T_c$ is a strong indication of an equal spin-pairing, spin-triplet state for which the susceptibility is expected not to change below $T_c$. Furthermore, the absence of temperature dependence in the superconducting state along both directions implies that the orientation of the equal spin-pairing state is not locked to a single crystal axis, rather it is free to rotate, indicating a minimal effect of spin-orbit coupling \cite{ohm96}.

 In principle, it is possible that strong spin-orbit scattering could suppress the temperature dependence of the spin shift \cite{and59} as $K_{sc}/K_{normal} = 1- 2l/\pi\xi_0$ for $l < \xi_0$ in the strong scattering limit, where $l$ is the electron mean free path and $\xi_0$ is the Ginzburg-Landau (GL) coherence length.  However, for a \upc\ sample with $\rho_0$ = 10 $\mu \Omega$cm, $l = 580$ \AA \cite{wu94}.  The upper critical field,  determined from the AC-susceptibility through NMR probe detuning as described below, gives $\xi_0 = 57$ \AA . Thus, we conclude that, since $\xi_0 \ll l$, our \upc\ sample is in the clean limit and spin-orbit scattering cannot account for the temperature independence of $K_s$ in the  SC state.
 
 \begin{figure}[t!]
\begin{center}
		\includegraphics[width=.45\textwidth]{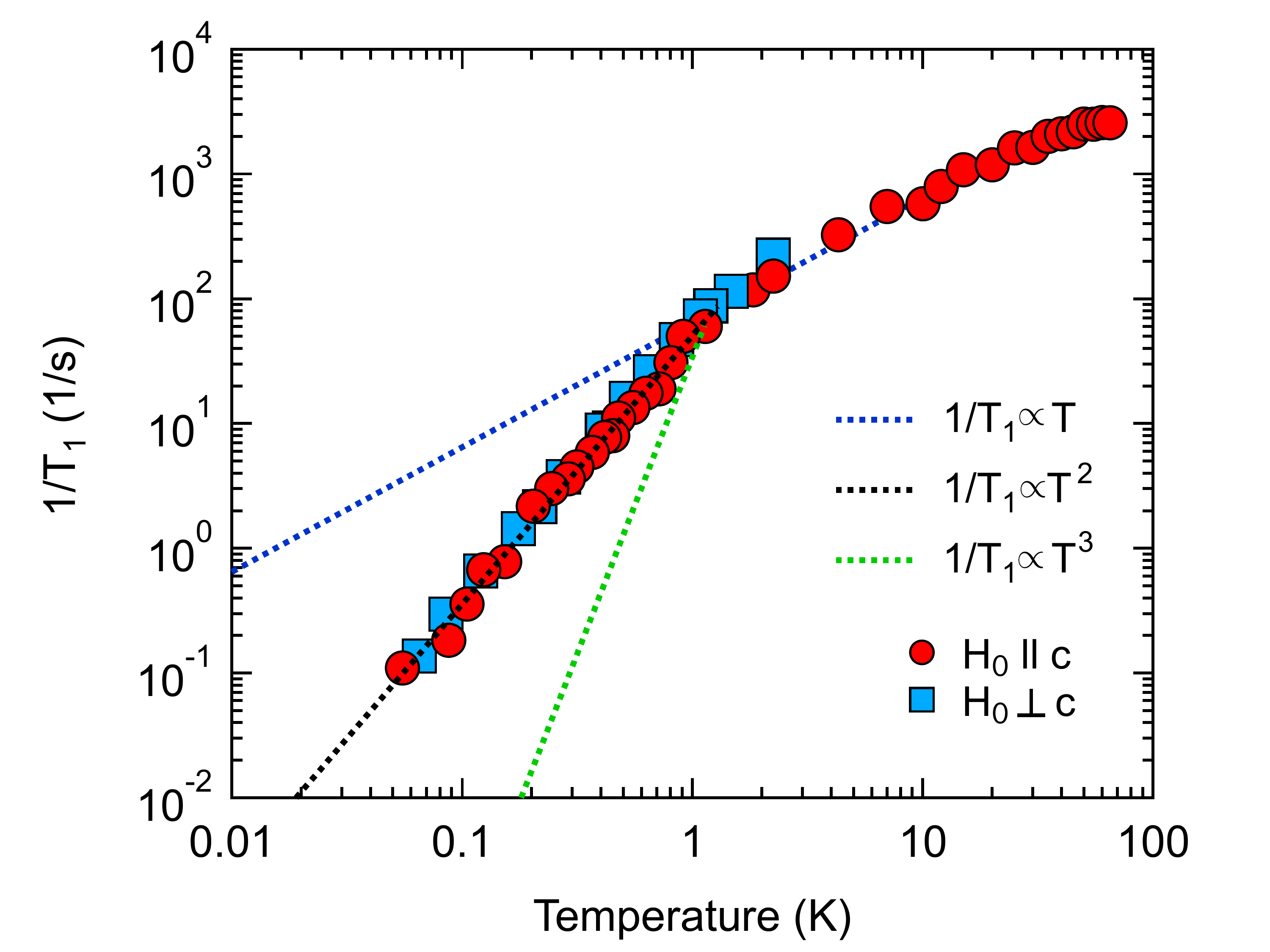}
		\caption{The spin lattice relaxation rate in the normal and superconducting states for \Hparac\ (red, circles), and in the superconducting state for \Hperpc\ (blue, squares). Dashed lines indicate power law temperature dependences.}
\label{t1}
\end{center}
\end{figure}

Now we turn to the spin-lattice relaxation rate measurements, $1/T_1$, which probe the imaginary component of the $q$-dependent dynamic susceptibility, $1/T_1\propto \sum_q \chi '' (q, \omega_0)$.  In the high temperature regime, $T > 40$ K,  for \Hparac , $1/T_1$ approaches a temperature independent value consistent with local moment behavior \cite{kur00}. At temperatures $T_c <  T < 40$ K, $1/T_1$ is linear with temperature, so called Korringa behavior, which is expected of a heavy Fermi liquid.  The modified Korringa law is given by $T_1TK_s^2 \equiv S \mathcal{K}(\alpha)^{-1}$, where $S$ is the Korringa constant defined as  $S = \hbar \gamma_e^2/ 4\pi \gamma_N^2 k_B$, and $\mathcal{K}(\alpha)$ is an enhancement factor which is equal to 1 for non-interacting electrons, greater than 1 for antiferromagnetic exchange correlations, and less than 1 for ferromagnetic correlations \cite{mor63}. The Korringa relation should taken as anisotropic, as evident from the anisotropy of the shifts, such that $T_{1,j}TK_{s,i}^2 \equiv S \mathcal{K}_i(\alpha)^{-1}$ where $i$ is parallel or perpendicular to the crystal $\hat c$ direction and $j$ are the directions perpendicular to $i$. For \upc, we find that $\mathcal{K}_c(\alpha) = 0.03$ and $\mathcal{K}_{ab}(\alpha)$ = 0.64  at $T = $ 2 K suggesting a strong ferromagnetic exchange enhancement dominant in the $\hat c$ direction. It is useful to compare the value for $\mathcal{K}_c(\alpha)$ with the intermetallic TiBe$_2$ for which $\mathcal{K}(\alpha) = 0.032$ \cite{tak84}, which is nearly ferromagnetic.

Additional evidence for ferromagnetic fluctuations come from the Wilson ratio $\mathcal{R} = \frac{\pi^2R}{3 C}\frac{\chi(0)}{\gamma}$ where $R$ is the ideal gas constant, $C$ is the Curie constant, $\chi(0)$ is the low temperature susceptibility, and $\gamma$ is the electronic specific heat coefficient. Using $\chi_c(T = 2 $K$) = 5.6$x$10^{-3}$ emu/mol U, $\chi_{ab}(T = 2 $K$) = 3.5$x$10^{-3}$  emu/mol U and $\gamma = 0.075$ J/mol U K$^2$ \cite{mei84}, we find that $\mathcal{R}_c \approx 3.3$, and  $\mathcal{R}_{ab} \approx 2$, both larger than the expected $\mathcal{R} = 1$ for a heavy Fermi liquid and characteristic of ferromagnetic enhancement of the electronic interactions.

\begin{figure}[b!]
\begin{center}
		\includegraphics[width=.45\textwidth]{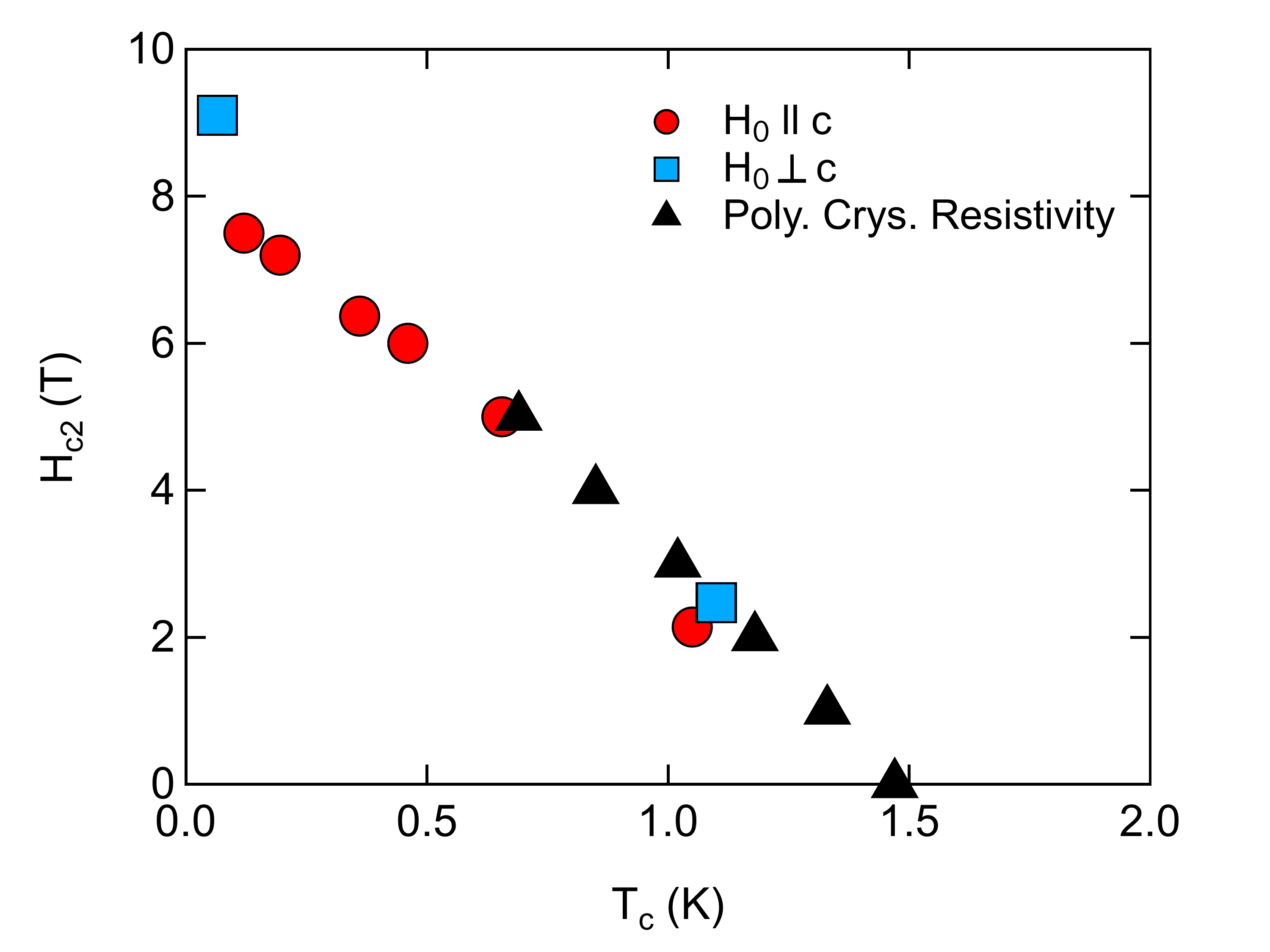}
		\caption{The upper critical field, $H_{c2}$ - superconducting transition temperature, $T_c$ phase diagram. The orientationally dependent $H_{c2}$ and $T_c$ were measured \textit{in situ} using the NMR coil and turning circuit (red, blue) while the polycrystalline $T_c$ were measured using resistivity(black, triangle).  }
\label{hc2}
\end{center}
\end{figure}

For $T < T_c$, the relaxation rate $1/T_1$ decreases as $1/T_1 \propto T^2$ along both orientations.  This is an unusual power law as $1/T_1$ generally decreases below $T_c$ as an exponential for a fully gapped Fermi surface, $T^3$ for a Fermi surface with line nodes, and $T^5$ for a Fermi surface with point nodes \cite{sig91}. We note that the magnetization recovery did not follow a simple exponential law in the SC state, rather a stretched exponential behavior indicating a distribution of relaxation times \cite{joh06}, likely due to misaligned crystallites and the vortex state. The $T^2$ behavior is indeterminate as to the orbital structure of the Cooper pairs, however the salient features are that $1/T_1$ decreases as a power law and there is no evidence of a Hebel-Slichter coherence peak \cite{heb59}.  With these factors considered, it is likely that \upc\ does not have a fully gapped Fermi surface, rather a nodal structure, the details of which may become more clear from measurements on single crystals.

The upper critical field, $H_{c2}$, and $T_c$ were measured on the aligned sample along both orientations by  AC susceptibility, and $T_c(H)$ was also determined on polycrystalline samples from resistivity measurements, Fig. \ref{hc2}. The upper critical field from  AC-susceptibility is $H_{c2}(T\to 0)$ = 7.8 (9.2) T for \Hparac\ (\Hperpc), although it should be noted that due the misaligned crystallites, evident from the spectra, the measured $H_{c2}$ could have errors of approximately 10 \%.  The initial slope of the upper critical field  found from the resistivity measurements is $dH_{c2}/dT_c\approx -10 $ T/K.  The zero-temperature orbital limiting field can be estimated in the weak coupling limit from the initial slope of $H_{c2}$ by the Werthamer-Helfand-Hohenberg relation as $H_{c2, orb} (0) = 0.693(-dH_{c2}/d T_{c}) T_c$ \cite{wer66, hak67},  which gives $H_{c2,orb}(0) = $ 10.2 T for the polycrystalline sample, consistent with the measured values in both orientations. Thus, the GL coherence length can be estimated as $\xi_0 \equiv \sqrt{\Phi_0/2\pi H_{c2,orb}}=57$ \AA , where $\Phi_0 = 2.07$x$10^{-15}$ Tm$^2$ is the flux quantum. 

The Pauli paramagnetic limiting field can be estimated from the Clogston-Chandrasekhar relation  \cite{clo62, cha62} for spin-singlet pairing {$H_{c2, p} (0)= 1.84 T_c$ in units of Tesla, which gives a value of $H_{c2, p} (0)= 2.7$ T, i.e. $H_{c2,p}<H_{c2,orb}\approx H_{c2}$.  Thus, Pauli limiting, generally associated with spin-singlet pairing, is absent because in \upc\ as there is no change in spin susceptibility as temperature decreases through $T_c$ and, consequently, the Pauli limiting field goes to infinity. This indicates that the upper critical field is dominated by orbital effects as additional supporting evidence of spin-triplet pairing.

In summary, for the first time $^{195}$Pt NMR and $H_{c2}$ have been measured on an aligned powder sample of the heavy-fermion superconductor \upc.  In the normal state, $K$ deviates from the bulk $\chi$ at a temperature $T^* = 25$ K, typical of heavy-fermion materials. $1/T_1$ goes from a localized $T$-independent behavior at high temperatures to a Korringa $T$-linear behavior at lower temperatures. The modified Korringa law shows that \upc\ is ferromagnetically exchange enhanced, consistent with an enhanced Wilson ratio. In the superconducting state, $K_s$ is temperature independent and unchanged relative to the normal state along both orientation, suggesting spin-triplet superconductivity with spin-orbit coupling too small to lock the equal spin pairs to the crystal lattice. Below $T_c$, $1/T_1$ follows a $T^2$ law likely indicative of a nodal gap structure. Further measurements on single crystals are necessary to clarify the orbital component of the superconducting pairs. From AC-susceptibility and resistivity we show that $H_{c2}$ is consistent with the orbitally limiting field and exceeds the spin-singlet Pauli limiting field as a further indication of spin-triplet superconducting Cooper pairs. Collectively, these data are compatible with \upc\ being in a class of spin-triplet superconductors that show a temperature independent $K_s$ below $T_c$ which includes UPt$_3$ \cite{tou96}, Sr$_2$RuO$_4$ \cite{ish98}, UNi$_2$Al$_3$\cite{ish02}, and UCoGe \cite{hat14}.
 
 \textbf{Acknowledgments}
 
We thank J. M. Lawrence for useful discussion.  Work at Los Alamos was performed under the auspices of the U.S. Department of Energy, Office of Basic Energy Sciences, Division of Materials Sciences and Engineering. A.M., G.K., and N.N. would also like to acknowledge postdoctoral fellowships funded by the Los Alamos LDRD program.  
 
\bibliography{u2ptc2bib}

\end{document}